\documentclass[twocolumn]{aastex631}

\begin{document}

\title{The S-PLUS 12-band photometry as a powerful tool for discovery and classification:\\ 
ten cataclysmic variables in a proof-of-concept study}

\correspondingauthor{Raimundo Lopes de Oliveira}
\email{raimundo.lopes@academico.ufs.br}

\author{Raimundo Lopes de Oliveira}
\affiliation{Departamento de F\'isica, Universidade Federal de Sergipe, Av. Marechal Rondon, S/N, 49100-000, S\~ao Crist\'ov\~ao, SE, Brazil}
\affiliation{Observat\'orio Nacional, Rua Gal. Jos\'e Cristino 77, 20921-400, Rio~de~Janeiro, RJ, Brazil}

\author{Amanda S. de Araújo}
\affiliation{Observat\'orio Nacional, Rua Gal. Jos\'e Cristino 77, 20921-400, Rio~de~Janeiro, RJ, Brazil}
\affiliation{Departamento de F\'isica, Universidade Federal de Sergipe, Av. Marechal Rondon, S/N, 49100-000, S\~ao Crist\'ov\~ao, SE, Brazil}

\author{Angela C. Krabbe}
\affiliation{Departamento de Astronomia, Instituto de Astronomia, Geof\'isica e Ci\^encias Atmosf\'ericas, Universidade de S\~ao Paulo, R. do Mat\~ao 1226, Cidade Universit\'aria, 05508-090, S\~ao Paulo, SP, Brazil}

\author{Claudia L. Mendes de Oliveira}
\affiliation{Departamento de Astronomia, Instituto de Astronomia, Geof\'isica e Ci\^encias Atmosf\'ericas, Universidade de S\~ao Paulo, R. do Mat\~ao 1226, Cidade Universit\'aria, 05508-090, S\~ao Paulo, SP, Brazil}

\author{Koji Mukai}
\affiliation{CRESST II and X-ray Astrophysics Laboratory, NASA/GSFC, Greenbelt, MD 20771, USA}
\affiliation{Department of Physics, University of Maryland, Baltimore County, 1000 Hilltop Circle, Baltimore, MD 21250, USA}

\author{Luis A. Guti\'{e}rrez-Soto}
\affiliation{Instituto de Astrof\'{i}sica de La Plata (CCT La Plata - CONICET - UNLP), B1900FWA, La Plata, Argentina}

\author{Antonio Kanaan}
\affiliation{Departamento de Física, Universidade Federal de Santa Catarina, Florianópolis, SC 88040-900, Brazil}

\author{Romualdo Eleutério}
\affiliation{Observat\'orio Nacional, Rua Gal. Jos\'e Cristino 77, 20921-400, Rio~de~Janeiro, RJ, Brazil}

\author{Marcelo Borges Fernandes}
\affiliation{Observat\'orio Nacional, Rua Gal. Jos\'e Cristino 77, 20921-400, Rio~de~Janeiro, RJ, Brazil}

\author{Fredi Quispe-Huaynasi}
\affiliation{Observat\'orio Nacional, Rua Gal. Jos\'e Cristino 77, 20921-400, Rio~de~Janeiro, RJ, Brazil}

\author{William Schoenell}
\affiliation{The Observatories of the Carnegie Institution for Science, 813 Santa Barbara St, Pasadena, CA 91101, USA}

\author{Tiago Ribeiro}
\affiliation{Rubin Observatory Project Office, 950 N. Cherry Ave., Tucson, AZ 85719, USA}

\begin{abstract}

Multi-band photometric surveys provide a straightforward way to discover and classify astrophysical objects systematically, enabling the study of a large number of targets at relatively low cost. Here we introduce an alternative approach to select Accreting White Dwarf (AWD) candidates following their spectral energy distribution, entirely supported by the twelve photometric bands of the Southern Photometric Local Universe Survey (S-PLUS). The method was validated with optical spectroscopic follow-up with the Gemini South telescope which unambiguously established ten systems as cataclysmic variables (CVs), alongside Swift X-ray observations of four of them. 
Among the ten CVs presented here are those that may be low-luminosity intermediate polars or WZ Sge-type dwarf novae with rare outbursts, two subclasses that can be easily missed in time-domain and X-ray surveys, the two methods currently dominating the discovery of new CVs.  Our approach based on S-PLUS provides an important, complementary tool to uncover the total population of CVs and the complete set of its subclasses, which is an important step towards a full understanding of close binary evolution, including the origin of magnetic fields in white dwarfs and the physics of accretion. Finally, we highlight the potential of S-PLUS beyond AWDs, serving other surveys in the characterization of their sources.

\end{abstract}


\section{Introduction}

As the end-product of $\sim$\,95\% of the stars in the Milky Way and under extreme physics conditions like high densities (up to 10$^9$ kg\,m$^{-3}$) and magnetic fields ($\sim$ 10$^{6}$-10$^{9}$\,Gauss) involving degenerate electrons, white dwarfs (WDs) serve as a laboratory for stellar structure and evolution \citep[see][for a recent review]{2022FrASS...9....6I}. 
Those that are paired with a companion close enough to efficiently transfer its material increase their relevance to our understanding, not only of their stellar parameters and evolution, but also of binary evolution and astrophysical plasma \citep[see][for a recent review]{2023arXiv230310055W}. Included in this context are the Accreting White Dwarfs (AWDs) in Cataclysmic Variables (CVs) and Symbiotic binary systems, in which the donor is usually a main sequence and a red giant star, respectively. There are many subclasses of CVs with their own set of characteristics,
including, for example, variability properties, absolute magnitudes and spectral-energy distributions (SEDs) in the optical, and properties in X-rays.

Three classical signatures of CVs allow us to reveal them: (i) the variability that gives the name to the class, including large outbursts and flickering; 
(ii) the presence of emission lines, some of them of high excitation; and 
(iii) thermal X-ray emission with moderate-to-low luminosity ($<$10$^{34}$\,erg\,s$^{-1}$ at 0.3-10 keV). 
Together, these features contrast with what is observed in isolated (non-accreting) WDs and single main sequence stars, providing evidence of ongoing accretion in binary systems. 

Many surveys have contributed to systematically revealing CVs, mostly supported by detecting variability in the optical and spectral hardness in X-rays. Examples of surveys based on optical photometry are the 
Palomar Transient Factory \citep[PTF;][]{2014ASPC..490..389M,10.1093/mnras/stu2105},
the Catalina Real-Time Transient Survey \citep[CRTS;][]{Drake_2009,10.1093/mnras/stu639},
the Optical Gravitational Lensing Experiment  \citep[OGLE; e.g.,][]{2013AcA....63..135M},
the All Sky Automated Survey \citep[ASAS;][]{2002AcA....52..397P}, 
and the GAIA Survey \citep{2016A&A...595A...1G}. 
In X-rays, substantial contributions emerged from the ROSAT All-Sky Survey \citep{1999A&A...349..389V}, Swift/BAT \citep{swiftbat} and INTEGRAL surveys \citep{Bird_2016}.
There have been surveys that used SEDs
that are not spectroscopic surveys in the usual sense, such as the first stage of Sloan Digital Sky Survey \citep[SDSS;][]{2002AJ....123..430S}, and older surveys such as the Palomar
Green survey \citep[see][]{1993PASP..105..805R} and Kiso UV Excess object survey \citep{1986ASSL..121..619M}, all based on filter
photometry.  Then there were some objective prism surveys such as the Case Western
survey \citep[for example,][]{1982PASP...94..682S} and the Hamburg Schmidt survey \citep{1995A&AS..111..195H}.  
Optical spectroscopy is usually used to follow-up preselected objects of interest.  The Large Sky Area Multi-Object fiber Spectroscopic Telescope is one of the exceptions of large surveys focused on optical spectroscopy \citep[LAMOST;][]{2012RAA....12..723Z}. The Vera C. Rubin Observatory \citep{2019ApJ...873..111I}, first referred to as Large Synoptic Survey Telescope (LSST), in the optical, promises to significantly decrease optical detection limits for AWDs, as eROSITA \citep{2021A&A...647A...1P} has demonstrated to be for low luminous X-ray systems \citep[e.g.,][]{2024A&A...686A.110S,2024arXiv240816053R}. The detection is naturally biased to brighter systems displaying outbursts serendipitously flagged by optical surveys, while X-ray detection favors subclasses with higher X-ray luminosities. Given the presence of a significant minority of CVs that are faint in X-rays and do not exhibit frequent outbursts (see Figure 4 in \citealt{2009MNRAS.397.2170G}), it is prudent to complement the LSST and eROSITA surveys with a SED-based investigation.  Such an effort would also provide an opportunity to discover unusual objects and/or subclasses that might alter our perception of CVs as a whole.

A growing number of studies point to the fact that extreme cases of AWDs, like those with fast-spinning WD, those with unusually low orbital periods, and those displaying low X-ray luminosities, or with a combination of such properties are underrepresented \citep[e.g.][and references]{2020ApJ...898L..40L}. What emerges from this is not only a matter of nature of the population but also the limitation imposed by the low number of representatives of different classes that can constrain the space of parameters necessary to better understand the formation, evolution, and properties of AWDs.

Here we present results from the proof-of-concept of an alternative approach to identifying CVs. Instead of relying on their variability (primarily in the optical) and X-ray properties, we identify H$\alpha$ emitters and investigate their spectral emission distribution from the 12-band photometric S-PLUS survey\footnote
{https://www.splus.iag.usp.br/} \citep{2019MNRAS.489..241M}. We demonstrate the effectiveness of this method in revealing CVs, by carrying out a spectroscopic follow-up of the candidates with the GMOS instrument at Gemini South observatory, alongside X-ray spectroscopy with the X-ray Telescope (XRT) at the Swift satellite. The first ten S-PLUS CV systems are presented (Table \ref{tab:systems}). 

\begin{table*}
	\centering
	\caption{Journal of systems.}
	\label{tab:systems}
	\begin{tabular}{lcccccccccc} 
		\hline
                  & \multicolumn{7}{c}{GAIA} && &S-PLUS \\
                  \cline{2-8} \cline{11-11}
		ID & R.A. & Dec. & dist$^{(2)}$ & G$^{(3)}$ & G$_{\rm abs}$$^{(4)}$ & G$_{\rm BP}$$^{(3)}$ & G$_{\rm RP}$$^{(3)}$ && A$_{V}$$^{(4)}$ & J0515\\
            & (J2000) & (J2000) & (pc) & (mag) & (mag) & (mag) & (mag) && (mag) & (mag) \\
  \hline

 S-PLUS CV1          &  04:11:35.29 & -30:22:14.6  & 612$_{-17}^{+16}$    & 17.04$\pm$0.03 &8.1		& 17.22$\pm$0.10 & 16.30$\pm$0.07 && 0.04$^{+0.10}_{-0.10}$ & 16.18$\pm$0.01 \\ 
 S-PLUS CV2          &  21:06:48.02 & -40:20:03.7  & 2095$^{+343}_{-259}$ & 16.41$\pm$0.01 &4.7		& 16.44$\pm$0.03 & 16.27$\pm$0.03 && 0.11$^{+0.10}_{-0.11}$ & 16.52$\pm$0.01 \\ 
 S-PLUS CV3$^{(1)}$  &  10:12:58.89 & -38:36:01.2  & 2157$^{+426}_{-367}$ & 17.52$\pm$0.01 &5.7		& 17.60$\pm$0.01 & 17.34$\pm$0.02 && 0.20$^{+0.12}_{-0.12}$ & 17.79$\pm$0.03 \\ 
 S-PLUS CV4$^{(1)}$  &  20:59:57.53 & -21:39:35.1  & 453$_{-26}^{+29}$    & 17.67$\pm$0.01 &9.3		& 17.76$\pm$0.03 & 17.41$\pm$0.03 && 0.15$^{+0.11}_{-0.11}$ & 18.38$\pm$0.03 \\ 
 S-PLUS CV5$^{(1)}$  &  10:41:52.78 & -20:28:22.6  & 1392$^{+766}_{-333}$ & 18.27$\pm$0.05 &7.5		& 18.32$\pm$0.15 & 18.05$\pm$0.14 && 0.10$^{+0.11}_{-0.10}$ & 17.77$\pm$0.02 \\ 
 S-PLUS CV6          &  04:52:31.13 & -44:11:04.2  & 652$_{-98}^{+156}$   & 19.58$\pm$0.03 &10.5	& 19.57$\pm$0.10 & 19.24$\pm$0.09 && 0.03$^{+0.10}_{-0.10}$ & 18.23$\pm$0.04 \\ 
 S-PLUS CV7$^{(1)}$  &  20:35:01.47 & -00:19:46.3  & 966$_{-129}^{+156}$  & 18.40$\pm$0.02 &8.2		& 18.45$\pm$0.05 & 17.93$\pm$0.05 && 0.25$^{+0.13}_{-0.11}$ & 18.83$\pm$0.02 \\ 
 S-PLUS CV8$^{(1)}$  &  10:17:59.70 & -36:05:13.8  & 1477$_{-409}^{+833}$ & 19.56$\pm$0.04 &8.5		& 19.90$\pm$0.14 & 18.70$\pm$0.11 && 0.22$^{+0.13}_{-0.12}$ & 19.57$\pm$0.09 \\ 
 S-PLUS CV9          &  21:16:39.81 & -23:43:16.9  & 369$^{+85}_{-54}$    & 19.14$\pm$0.01 &11.3	& 19.22$\pm$0.06 & 19.00$\pm$0.07 && 0.05$^{+0.10}_{-0.10}$ & 19.16$\pm$0.07 \\ 
 S-PLUS CV10         &  13:36:10.78 & -13:12:17.2  & 781$_{-179}^{+462}$  & 19.97$\pm$0.02 &10.4	& 20.41$\pm$0.14 & 18.99$\pm$0.06 && 0.16$^{+0.11}_{-0.11}$ & 19.87$\pm$0.08 \\ 
            \hline
	\end{tabular}
 \\
 Notes: (1) After pinpointing the systems from S-PLUS data, we noticed that they were pointed out as CV (or candidate to CV) in the AAVSO/VSX (https://www.aavso.org/vsx/) -- see text; (2) Geometric distances derived by \citet{2021AJ....161..147B} from GAIA astrometry; (3) Mean Gaia magnitudes from GAIA/EDR3 \citep{2021A&A...649A...1G}; (4) See Section \ref{sct:GAIA}
\end{table*}

\section{Observational data and databases}

\subsection{S-PLUS}
\label{sct:splus}

The Southern Photometric Local Universe Survey (S-PLUS) is on the way to observing $\sim$\,9300 deg$^2$ of the Southern celestial hemisphere in twelve bands \citep{2019MNRAS.489..241M}. 
This is being carried out with the T80-South, an 83\,cm telescope at the Cerro Tololo Inter-American Observatory, in Chile. Here we made use of its third data release (DR3), which covered about 1800 deg$^2$, exploring the aperture photometry from its five $ugriz$ broad-band filters and seven narrow-band filters to reveal CVs. 

The narrow filters of the S-PLUS are centered on the Balmer jump/[OII] (a.k.a. J0378), Ca H + K (J0395), H$\delta$ (J0410), G band (J0430), Mg\,b triplet (J0515), H$\alpha$ (J0660), and the Ca triplet (J0861) \citep[see Table 2 in][]{2019MNRAS.489..241M}. The filters J0395, J0410, and J0430 cover the H$\epsilon$, H$\delta$, and H$\gamma$ lines, respectively. However, their magnitudes cannot be directly applied to compare the intensities of those Hydrogen lines given that their widths ($\Delta \lambda$) are large enough (103\,\AA\ for J0395 and 201\,\AA\ for the other two bands) to include a significant contribution of the continuum. Nevertheless, the use of such filters is valuable to verify a key point of this project, which is the bluishness of the S-PLUS SED due to the presence of such lines and the white dwarf emission. Magnitudes of the filter J0660 ($\Delta \lambda$ $\sim$ 147\,\AA, centered at 6614\,\AA) were used as a proxy of the H$\alpha$ line, by comparing them with the mean magnitudes from the contiguous $r$ and $i$ filters. From the broad-band filters, we have the coverage at the blue end with the $u$ filter, and the red part of the spectrum with the $r$, $i$, and $z$ filters. The details of the procedures exploring the S-PLUS photometry in this work will be summarized in Section \ref{sect:met}.

To minimize instrumental artifacts and use only well-determined photometric values, our working sample was composed of objects whose photometry met two conditions: sources detected in all twelve bands and uncertainties below 0.1 magnitudes in each one of the filters. Although it compromises completeness, such an approach favors sample purity, which is the guideline of this work.

\subsection{GAIA}
\label{sct:GAIA}

The angular distance between the S-PLUS and GAIA coordinates is less than 0.25 arcsec for all targets. Their correspondence was double-checked in the S-PLUS image discarding confusion with nearby sources having magnitude consistent with those from GAIA. We adopted geometric distances derived from Gaia parallax measurements as reported by \citet{2021AJ....161..147B}. These values were useful to exclude systems in our first target selection with large, unrealistic distances beyond the capability of GAIA, conservatively serving as a second criterion to exclude potential QSO systems from our initial sample -- we also adopted the flag of potential QSO available in the S-PLUS catalog \citep[after][]{2021MNRAS.507.5847N}. This is especially important because QSOs mimic some features of CVs as seen from photometric SEDs. 
As for our targets (Table \ref{tab:systems}), the GAIA Renormalised Unit Weight Error (RUWE)
parameter values are restricted to the range 0.939--1.062, indicative of a good astrometric solution and thus a reliable distance determination.
Finally, distances were used to estimate the X-ray luminosity of the four targets observed with the Swift telescope in response to our request. 

For the absolute G magnitudes, we subtract from the apparent magnitudes the corresponding values of 5$\times$log($<$best estimate distance$>$/$<$10 pc$>$)
and the A$_G$, estimated using the extinction map of \citet{2024arXiv240303127D}\footnote{http://astro.uni-tuebingen.de/nh3d/nhtool} and using A$_G$/A$_V$=2.8/3.1=0.90.
All 10 objects of our sample have small extinctions, the largest being A$_V$\,=\,0.25$^{+0.13}_{-0.11}$ for S-PLUS CV7 (see Table \ref{tab:systems}). The uncertainties in the absolute G magnitudes are dominated by uncertainties
in the distance estimates. Values are presented in Table \ref{tab:systems}.

\begin{table*}
	\centering
	\caption{Journal of Gemini and Swift observations.}
	\label{tab:obs}
	\begin{tabular}{cccccccc}
		\hline
             & \multicolumn{4}{c}{Gemini}                                                      && \multicolumn{2}{c}{Swift}\\
\cline{2-5} \cline{7-8}

ID           &  Date        &  \multicolumn{2}{c}{Central wavel. / exp. times}   &  Spectral coverage && Date & XRT exp. \\  
\cline{3-4} \cline{7-8}
             &  (UT)        & (\AA) / (s)              &  (\AA) / (s)            & (\AA)       && (UT)       & (s)    \\
 
\hline
S-PLUS CV1   &  2021-10-11  &  5200  /  2$\times$500   &  5250  /  2$\times$500  &  3664--6866 && 2023-04-10 & 4896   \\
S-PLUS CV2   &  2023-06-11  &  5100  /  2$\times$500   &  5700  /  2$\times$500  &  3558--7302 && ...        & ...    \\
S-PLUS CV3   &  2023-05-01  &  5100  /  2$\times$1000  &  5700  /  2$\times$1000 &  3558--7302 && ...        & ...    \\
S-PLUS CV4   &   2023-05-30  &  5100  /  2$\times$1000  &  ... & 3558--6686 && 2023-04-10 & 4111   \\
             &   2023-05-31  &  ...  &   5700  /  2$\times$1000 &   4132--7301 &&  &    \\
S-PLUS CV5   &  2023-06-14  &  5100  /  2$\times$1000  &  5700  /  2$\times$1000 &  3559--7303 && ...        & ...    \\
S-PLUS CV6   &  2021-10-11  &  5200  /  2$\times$1200  &  5250  /  2$\times$1200 &  3664--6866 && 2023-04-13 & 3801   \\
S-PLUS CV7   &  2023-06-11  &  5100  /  2$\times$1200  &  5700  /  2$\times$1200 &  3559--7302 && ...        & ...    \\
S-PLUS CV8   &  2023-05-27  &  5100  /  2$\times$1200  &  ... &  3559--6701 && ...        & ...    \\
             &  2023-06-13  &  5100  /  1$\times$1200  &  5700  /  3$\times$1200 &  3559--7303 && ...        & ...    \\
S-PLUS CV9   &  2023-06-14  &  5100  /  3$\times$1200  &  5700  /  1$\times$1200 &  3559--7303 && 2023-04-09 & 4429   \\
&  2023-06-29  &  ...  &  5700  /  2$\times$1200 &  4132--7303 &&   &    \\
S-PLUS CV10  &  2023-06-11  &  5100  /  5$\times$1200  &  5700  /  4$\times$1200 &  3558--7302 && ...        & ...    \\
\hline
	\end{tabular}
\end{table*}

\subsection{Gemini South}
\label{sct:GS}

Long-slit spectroscopy of targets of our exploratory sample was carried out with the Gemini Multi-Object Spectrograph (GMOS) coupled to the 8\,m telescope at the Gemini South Observatory. The follow-up was approved for the GS-2023A-Q-411 and GS-2021B-Q-407 programs in poor weather queue, designated as ``Band 4'', and GS-2023A-FT-212 in the Fast Turnaround observing mode (PI in all proposals: R. Lopes de Oliveira). The spectra covered the wavelength range from about 3560\,\AA\ to 7300\,\AA\ in two settings with the B600 grating, one favoring the blue and the other the red part of the optical light with slight differences in the wavelength coverage from target to target. The individual spectra for each source were combined into the final one. S-PLUS CV8 and CV9 have spectra obtained on different days  ($\sim$\,15-day difference)but after checking we found no significant changes so they were combined as well. However, the noise dominates the blue end in such a way that the spectra were only useful from above 3800\,\AA\ -- enough for our purposes. We used a slit width of 1 arcsec. The spectral resolution was about 5.6\,\AA. Details of the Gemini observations are presented in Table \ref{tab:obs}.

Data reduction was performed using the DRAGONS Data Reduction Software V3.1.0 \citep{2023RNAAS...7..214L}, following standard procedures. Stellar spectra for flux calibration were those available weekly from the observatory and therefore did not correspond to the (bad) weather conditions at the time of the observations or even the air mass associated with scientific objects. The practical result of this is that we can compare the relative fluxes along a given spectrum for a given grating configuration -- assuming the atmospheric contribution does not vary during the exposures and its effects do not depend on wavelength --  but the spectra do not have an absolute flux calibration. This restriction was adopted to be less restrictive of the Gemini telescope time because relative calibration is enough for our purposes.

\subsection{Swift}

X-ray follow-up of the four closest targets ($d < 700$ pc, to increase the chance of detection with the snapshots) was carried out using the X-ray Telescope (XRT) and the Ultra-violet Optical Telescope (UVOT) cameras on board the Neil Gehrels Swift Observatory as targets of opportunity (PI: Lopes de Oliveira). In all instances, XRT operated in the Photon Counting mode, while the UVOT, set to the ``filter of the day'', varied for each target. Here we explore only the XRT data (see Table \ref{tab:obs}), because the UVOT observations were not simultaneous with the ground-based observations and they are not relevant in the context of this paper. Data reduction followed the standard procedures with the HEASoft/NASA\footnote{https://heasarc.gsfc.nasa.gov/docs/software/lheasoft/} V6.30.1 and using the  HEASARC's calibration database (CALDB) 2022033. Spectral investigation made use of the \textsc{XSPEC} V12.12.1 \citep{1996ASPC..101...17A}.

The Swift snapshots (with durations $<$\,5000\,s) were aimed at providing initial insights into the systems in X-rays: confirming their X-ray emission, and assessing their hardness and luminosities. Three out of the four target systems were positively detected in X-rays and,  although the signal was poor, the data allowed us to construct spectra. For the only non-detection, we estimated an upper limit of its X-ray luminosity. To roughly quantify the X-ray properties with the available data, we adopted a single optically thin, collisionally-ionized diffuse plasma model, the \textsc{apec}, with the \textsc{tbabs} model accounting for the photoelectric absorption: \textsc{tbabs}$*$\textsc{apec}. The abundance table adopted was that of \citet{2009ARA&A..47..481A}. In all cases, the energy channels were grouped to ensure at least one count per bin but spectra were plotted such that the bins were combined to achieve a significance of at least 2 sigma limited to a combination of no more than 100 of them -- \textsc{setplot rebin 2 100} in XSPEC. C-Statistic was used as the fit statistic and Chi-Squared as the test statistic in X-ray spectral analysis with \textsc{XSPEC}. The luminosity values presented in the text were estimated assuming the distances presented in \citet{2021AJ....161..147B} (see Table \ref{tab:systems}).

\section{Methodology}
\label{sect:met}

Our exploration of the S-PLUS data prioritized purity over completeness by simultaneously looking for common characteristics of three emission sites in AWDs: (i) the material undergoing accretion, (ii) the accreting white dwarf, and (iii) the donor star. In the following, we discuss these features and the foundations of this work to uncover AWDs.

The accreting, partially ionized matter, when irradiated by the accreting white dwarf itself, leads to the production of emission lines. In the optical, the strongest is the H$\alpha$ line, sometimes it is the H$\beta$ line under the situation in which the medium is optically thick to H$\alpha$ photons \citep[e.g.,][]{1977ApJ...217..815S}. Other lines of the Hydrogen Balmer series may be present in AWDs as well. Transitions from He are commonly observed, as those of He\,I at 4026, 4471, 4921, 5015, 5876, 6678, and 7065\,\AA, and He\,II at 4686\,\AA. As for the other lines, He\,II\,4686\,\AA\ is observed in emission in a variety of AWDs and its presence is not conclusive to classify a system as belonging to one class or another. However, a higher strength of the (high excitation) He\,II\,4686\,\AA\ line compared to, for example, the strength of the H$\beta$ line, is suggestive of magnetic systems \citep[e.g.,][]{1992PhDT.......119S}. We begin by highlighting probable H$\alpha$ emitters from S-PLUS data.

The H$\alpha$ line is a classic indicator of accretion not only in AWD systems but in others as well and is the first signature explored in this work. Evidence of H$\alpha$ line in emission can be found in the S-PLUS data from observations with the J0660 filter \citep[see Fig. 6 in][]{2019MNRAS.489..241M}. However, the absence of a narrow filter covering the continuum around H$\alpha$ leads us to have to compare the magnitude of the J0660 filter to those of broad-band filters to identify potential H$\alpha$ emitters. {\it Criterion 1:} we adopted the average of the $r$ and $i$ magnitudes as a proxy for the continuum around H$\alpha$, such that sources with [($r$-$i$)/2 - mag$_{J0660}$] greater than 0.1 magnitudes in the S-PLUS were flagged as H$\alpha$ emitters and made up our initial sample. 

The second characteristic we use to identify AWDs is their typical blue colors for shorter wavelengths. This is due to different components. The first one is the (hot) white dwarf surface, which emits prominently in the UV region, leading to a significant continuum that increases towards the blue end -- which is the most crucial feature for identifying white dwarfs in general. If an accretion disk is present, a fraction of the blue continuum emerges from its hot inner regions as well. Another component, if present, is that from emission lines below 5,000 \AA\, -- especially ones already noted at the beginning of this section. While the individual presence of these lines cannot be determined from S-PLUS data (see Section \ref{sct:splus}), the overall blue emission from them, as well as from the white dwarf and the tentative accretion disk, can be observed. {\it Criterion 2:} we searched for a blue emission as expected from AWDs by considering sources with mag$_{J0395}$/mag$_{J0515}$ $<$ 1.

\begin{figure}[t!]
 \includegraphics[width=9.5cm]{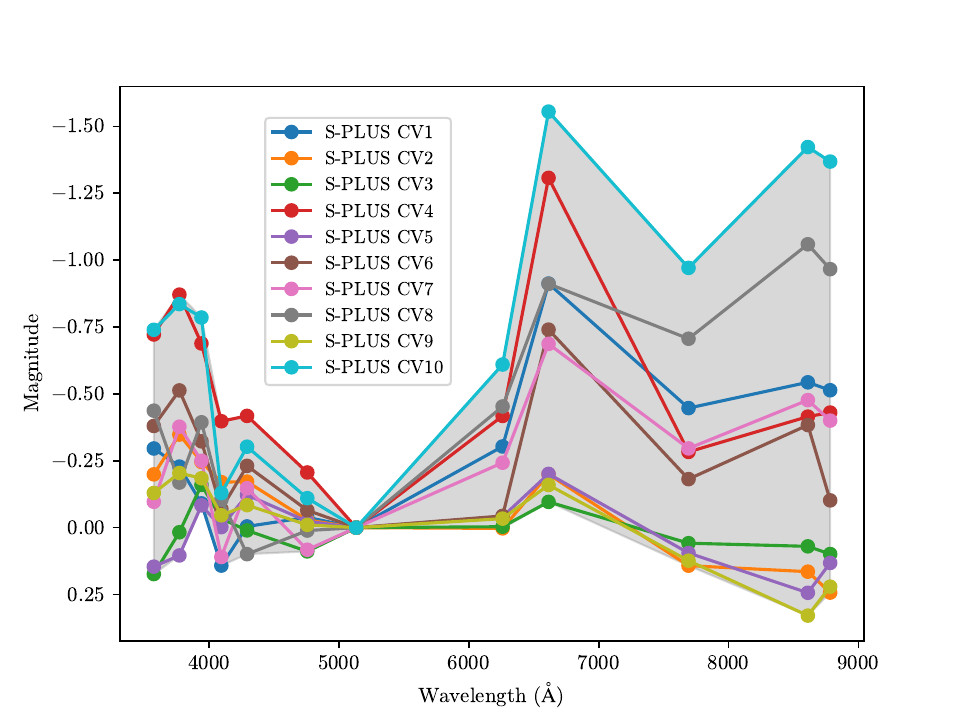}

   \caption{Spectral energy distribution from the S-PLUS photometry}
    \label{fig:splussed}
\end{figure}

The third aspect considered was the red emission, serving as an indication of the presence of the donor star in the system. Its existence disrupts the continuous decline of the WD contribution towards longer wavelengths. At times, it generates such an excessive amount in the red spectrum that a V-shaped SED becomes evident in the interplay between the blue and the red ends of the SED in the optical: {\it this is the gold standard sought for identifying H$\alpha$ emitters with a higher likelihood of being a CV}. {\it Criterion 3:} at this stage, we designate the J0515 and $r$ magnitudes, with mag$_r$/mag$_{J0515}$ $<$ 1 as the third selection criterion to point out CV candidates. 

\begin{figure*}[t!]
 \includegraphics[trim= 20cm 4cm 0cm 0cm,clip,width=20.5cm]{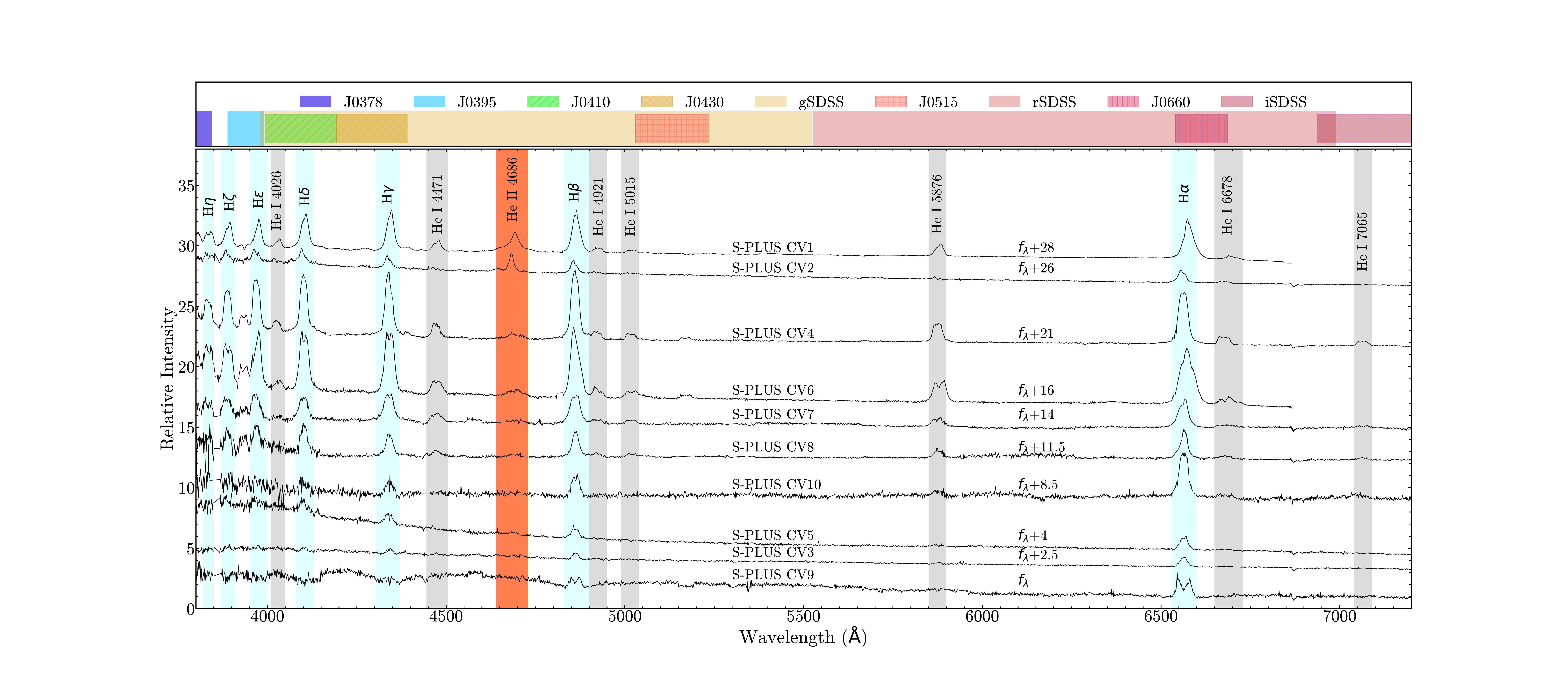}
 
   \caption{Optical spectra from GMOS/Gemini.}
    \label{fig:geminispect}
\end{figure*}

Finally, we proceed with a visual inspection of SEDs of S-PLUS sources that have met the three criteria outlined above (see Fig. \ref{fig:splussed}). We focused on the thirteen systems with clear detection of the expected features to proceed with optical spectroscopic follow-up with Gemini: eleven out of them were proven to be AWDs. These numbers give us a success rate of at least 85\% (11 out of 13 systems). However, the two ``outliers'', with optical spectra distinct from what is expected for AWDs, have S-PLUS SEDs that are still consistent with those of CVs; we suspect they were observed in a low activity state and plan to obtain new optical spectroscopy of them. Here we present ten out of the eleven proven AWDs (Table \ref{tab:systems}) - with the eleventh deserving a separate paper with a more in-depth investigation as it is already possible with the available data (Amanda S. de Araujo, in preparation). Their S-PLUS SEDs and optical spectra are shown in Fig. \ref{fig:splussed} and Fig. \ref{fig:geminispect}, respectively. To facilitate comparison, the magnitudes of a specific system in Fig. \ref{fig:splussed} were subtracted from the magnitude assigned to it in the filter J0515 ($\lambda_{eff}$\,=\,5133\,\AA; see magnitudes in Table \ref{tab:systems}) and error bars were omitted (the uncertainties are below 0.1 magnitudes in all cases). 
The fluxes of the optical spectra in Figure \ref{fig:geminispect} are divided by their values at 6500\,\AA\ and shifted by arbitrary constant values for clarity. Despite the fact that they lack absolute flux calibration (Section \ref{sct:GS}), this approach preserves the relative intensity and therefore the shape of the individual spectra.

While the adopted conservative approach compromises completeness, it ensures a purer sample. The primary goal at this initial, proof-of-concept stage is to validate the methodology and pave the way to increase the number of known AWDs from S-PLUS, and prompt further in-depth investigations. We have left it as the next step to explore fuzzy boundaries towards the limits accessible with S-PLUS and also do cross-correlation of its catalog with public catalogs of other missions.

\section{Results}

The following presents the systems in three sets according to their common features. While their S-PLUS SEDs and optical Gemini spectra are shown in Figures \ref{fig:splussed} and \ref{fig:geminispect}, respectively, the equivalent widths of the measurable H$\alpha$, H$\beta$, and He\,II\,4686 lines are presented in Table \ref{tab:ew}. 

\subsection{S-PLUS CV1 and S-PLUS CV2}
 
Emission lines of the Balmer series of Hydrogen were observed in the optical spectra of both S-PLUS CV1 and S-PLUS CV2, with larger equivalent widths for those of S-PLUS CV1 (Table \ref{tab:ew}). However, a notable difference between them is that the He\,I lines are stronger in S-PLUS CV1 than in S-PLUS CV2. One remarkable feature common to both systems is the He\,II\,4686\,\AA\ line, which in the case of S-PLUS CV2 has an equivalent width that matches that of the H$\beta$ line. 
In both systems, there is evidence of complex, likely multi-component emission lines -- asymmetric
profiles of H lines and, for S-PLUS CV1, some HeI lines may well be
double-peaked.

S-PLUS CV1 was positively detected with Swift/XRT but the available data only allows for constructing a noisy spectrum (see Fig. \ref{fig:swiftspect}). The \textsc{tabs}*\textsc{apec} model indicates a plasma with temperature ($k$T) of 4$^{+6}_{-3}$\,keV undergoing photoelectric absorption corresponding to that of a Hydrogen column of 10$^{22-23}$\,cm$^{-2}$. The unabsorbed flux is estimated to be 4$\times$10$^{-12}$ erg\,cm$^{-2}$\,s$^{-1}$ (0.3-10\,keV), which, at a distance of 612 pc, indicates a luminosity of 10$^{32}$ erg\,s$^{-1}$. Such an X-ray luminosity is unusually high for a non-magnetic CV, suggesting the possibility that this is a low-luminosity intermediate polar (LLIP), which is also supported by the relatively strong He\,II\,4686 line (Fig. \ref{fig:geminispect}).

While the He\,II\,4686/H$\beta$ ratio suggests the possibility that S-PLUS CV2 might be an AM Her type system, this is unlikely given that its absolute G magnitude (4.7) is far too bright to be a good AM Her candidate \citep[see Figure 2 of][] {2020MNRAS.492L..40A}.  Also, AM Her type systems tend to have strong HeI lines (such as the one in 5876\AA). 
Its absolute magnitude suggests the possibility of CV2 being
a nova-like system (i.e., a disk-dominated non-magnetic CV with a
high accretion rate).  High-inclination nova-like systems are known
to exhibit emission (rather than absorption) lines including HeII\,4686
at a level comparable to that seen in S-PLUS CV2 \citep[see, e.g.,][]{1994MNRAS.267..153H}.
Following the coordinates (with a Galactic latitude $b$\,=\,-42) and distance (2095 pc), S-PLUS CV2 appears to be in the Galactic halo.

\subsection{S-PLUS CV4, CV6, CV7, and CV8}

The systems S-PLUS CV4, CV6, CV7, and CV8 have similar optical spectra, in which the Balmer series of the Hydrogen, HeI, and HeII lines were observed. S-PLUS CV4 displays a flat-topped profile for H$\alpha$, and some HeI lines. Higher order Balmer lines (H$\gamma$ and so on) of S-PLUS CV6 and CV7 do look double-peaked.

As in the case of S-PLUS CV1 and CV2, they were caught in accretion-active phases.   However, contrary to S-PLUS CV1 and CV2, the emission in He\,II\,4686 of the four systems is modest, making their spectra similar to those of many dwarf novae but also similar to those of LLIPs.  
As described in the next paragraph, these two interpretations are fully consistent with the X-ray emission of two out of them, S-PLUS CV4 and CV6. 

It is worth noting that CV4 is listed on ASAS-SN variable star catalog as CV (ASASSN-V J205957.53-213935.2), CV7 is  SDSS J203501.46-001946.1 = ASASSN-14hb, listed as 18abssrbs  in Table 2 (CV candidates) of \citet{2020AJ....159..198S} and classified as a dwarf nova by \citet{2021ApJ...910..120K}, and S-PLUS CV8 is a transient source in the optical classified as a CV candidate (ASASSN-19eh) by  \citet{2014ApJ...788...48S}.

\begin{figure}[t!]
 \includegraphics[width=8.6cm]{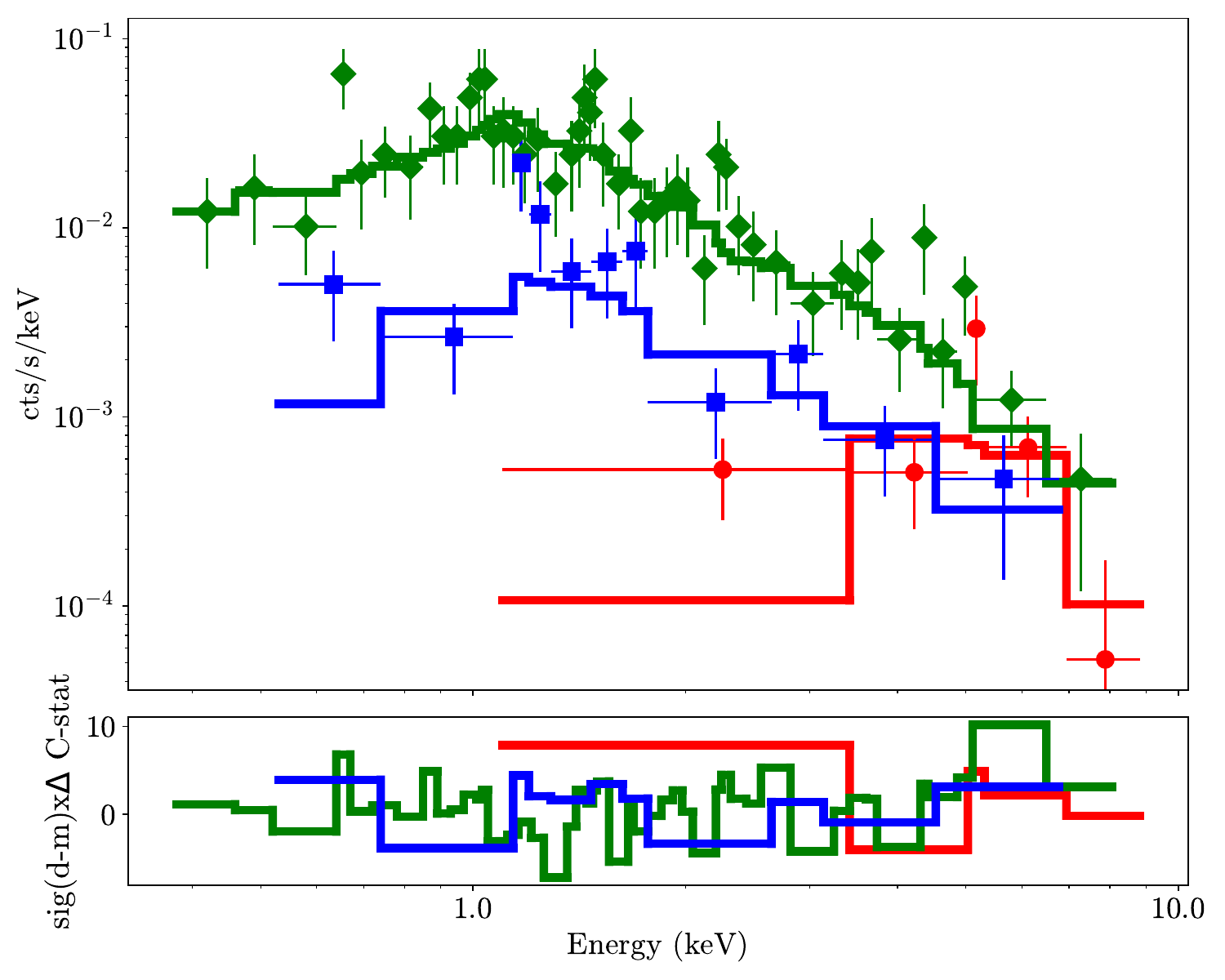}
    \caption{X-ray spectra from Swift/XRT: red circles for S-PLUS CV1, green diamonds for S-PLUS CV4, and blue squares for S-PLUS CV6.}
    \label{fig:swiftspect}
\end{figure}

S-PLUS CV4 is a ROSAT All-sky Survey \citep[RASS;][]{1999A&A...349..389V} source, while previous X-ray detections of S-PLUS CV6 include RASS, XMM-{\it Newton} Slew Survey \citep{2008A&A...480..611S}, and eROSITA \citep{2024A&A...682A..34M}. We observed both with Swift/XRT.
The signal accumulated for S-PLUS CV4 was good enough for a satisfactory spectral fit (Fig. \ref{fig:swiftspect}). From the \textsc{tabs}*\textsc{apec} model, we inferred the absorption to the equivalent in Hydrogen column (N$_{\rm H}$) of 1.2$^{+0.5}_{-0.5}$$\times$10$^{21}$\,cm$^{-2}$, a plasma temperature ($k$T) of 4.4$^{+2.0}_{-1.0}$\,keV, and unabsorbed flux of 3$\times$10$^{-12}$ erg\,cm$^{-2}$\,s$^{-1}$ at 0.3-10\,keV, so in a luminosity of 7$\times$10$^{31}$\,erg\,s$^{-1}$ for a distance of 453\,pc ($\chi^2$/d.o.f\,=\,133.83/169). 
The X-ray spectrum of S-PLUS CV6 was noisy (Fig. \ref{fig:swiftspect}). The same model as for S-PLUS CV4 applied in the S-PLUS CV6 spectrum gave a photoelectric absorption N$_{\rm H}$ = 4.2$_{-3.4}^{+4.3}$\,$\times$\,10$^{21}$\,cm$^{-2}$ but does not constrain the plasma temperature. Fixing the N$_{\rm H}$ parameter to 4.2\,$\times$\,10$^{21}$\,cm$^{-2}$, $k$T is constrained to 6$^{+43}_{-3}$\,keV. The unabsorbed flux in 0.3-10\,keV is estimated to be 7$\times$10$^{-13}$ erg\,cm$^{-2}$\,s$^{-1}$, so the luminosity, for a distance of 652\,pc, is 3.6$\times$10$^{31}$\,erg\,s$^{-1}$ ($\chi^2$/d.o.f\,=\,30.15/42). Thus, both S-PLUS CV4 and CV6 were modest hard and low-luminosity X-ray emitters at the time of their observations.

\subsection{S-PLUS CV3, CV5, CV9, and CV10}

The optical spectra of S-PLUS CV3, CV5, CV9, and CV10 revealed only the first Balmer series lines of Hydrogen, with no evidence of HeI and HeII. The double-peaked H$\alpha$ line is suggestive of an edge-on orientation to the systems S-PLUS CV9.  
S-PLUS CV9 and CV10 have faint enough absolute magnitudes to be consistent with a WD-photosphere-dominated quiescent dwarf nova. 
In contrast, S-PLUS CV3 and CV5 are too bright to fit this classification and must be accretion disk-dominated systems. 

S-PLUS CV3  has been classified by \citet{2015MNRAS.446.2251T} as an RR Lyr type star with a period of 0.50867 d (SSS J101253.5-383732). Given the Gemini spectrum, we classify it as a CV. In this case, the observed periodicity may represent its orbital period, which is unusually long for a CV. \citet{1987MNRAS.227...23W} proposed a relation, V$_{\rm abs}$(min)=9.72-0.337$\times$P$_{\rm orb}$(hr), as the absolute magnitudes of dwarf novae in quiescence.  If CV3 is a dwarf nova whose orbital period is 0.50867d (=12.2 hr), the extrapolation of this relationship predicts a quiescent V$_{\rm abs}$ of 5.6, compared to the observed G$_{\rm abs}$ of 5.7. 
S-PLUS CV5 is indicated as a CV from Zwicky Transient Facility photometry \citep[ZTF19aachvbn;][]{2021AJ....161..242F}.

The Swift/XRT observation failed to detect 
S-PLUS CV9, from which we estimate a conservative upper limit to the flux at 0.3-10 keV of (2-3)$\times$10$^{-13}$\,erg\,s$^{-1}$ from the 0.00542 counts per second estimated to the background and adopting a plasma emission with $k$T of 3.06 keV and 27.96 keV, respectively, in the web version of the Portable, Interactive Multi-Mission Simulator (PIMMS)\footnote{https://heasarc.gsfc.nasa.gov/docs/software/tools/pimms.html}. That is, the 0.3--10 keV flux does not depend very much on the assumed plasma temperature. These values imply in a luminosity up to 3-5$\times$\,10$^{30}$ erg\,s$^{-1}$ (to $d$\,=\,369 pc).

\begin{table}
	\centering
	\caption{Equivalent widths of optical emission lines.}
	\label{tab:ew}
	\begin{tabular}{cccc} 
		\hline
   & \multicolumn{3}{c}{EW (\AA)}           \\
\cline{2-4}   
ID & H$\alpha$ & H$\beta$ & He\,II\,4686 \\
\hline

S-PLUS CV1   & $-$107$\pm$4   & $-$68$\pm$3  & $-$33$\pm$3  \\
S-PLUS CV2   & $-$27$\pm$4    & $-$13$\pm$2  & $-$15$\pm$3  \\
S-PLUS CV3   & $-$19$\pm$4    & $-$6$\pm$2   & ...	    \\
S-PLUS CV4   & $-$166$\pm$6   & $-$110$\pm$3 & $-$17$\pm$5  \\
S-PLUS CV5   & $-$31$\pm$5    & $-$12$\pm$3  & $-$5$\pm$4   \\
S-PLUS CV6   & $-$213$\pm$5   & $-$123$\pm$3 & $-$15$\pm$3  \\
S-PLUS CV7   & $-$76$\pm$6    & $-$70$\pm$5  & $-$6$\pm$5   \\
S-PLUS CV8   & $-$68$\pm$14   & $-$48$\pm$11 & $-$8$\pm$14  \\
S-PLUS CV9   & $-$61$\pm$10   & $-$11$\pm$4  & ...	    \\
S-PLUS CV10  & $-$125$\pm$22  & $-$46$\pm$19 & ...	    \\

\hline
	\end{tabular}
\end{table}

\section{Final Remarks}

\subsection{The S-PLUS in the context of AWDs}

Through this work, we present 10 CVs, 5 of which have no publications on them to the best of our knowledge, while the other 5 are poorly studied variables (3 pointed out as CVs, 1 suspected CV, and 1 classified as an RR Lyr). This demonstrates the power of the S-PLUS
survey and validates our procedure for identifying CVs based on S-PLUS colors and SEDs.
In particular, the fact that half of our sample was previously unknown
shows that S-PLUS is competitive with the current generation of time-domain
surveys and X-ray surveys, and the effort to study the total population
of CVs will remain incomplete without a SED-based survey.

This is the case even when a CV is detected as an X-ray source.
S-PLUS CV4 and S-PLUS CV6 are both known X-ray sources from the days of
ROSAT all-sky survey, yet had not been identified as a CV until this work.
Notably, these objects had not revealed themselves in time-domain
surveys, suggesting that their variability is not so prominent as to
be picked up as transients and/or rare.

S-PLUS CV1, S-PLUS CV4, and S-PLUS CV6 are relatively faint in the optical, with absolute G magnitudes of 8.1, 9.3, and 10.5, respectively, so they are too faint to be nova-likes. While their brightness is in the right range to be the quiescent dwarf novae, there are no known outbursts of these systems. They have detectable HeII 4686 lines in emission and are relatively luminous in X-rays
(1$\times$10$^{32}$, 7$\times$10$^{31}$, and 3.6$\times$10$^{31}$ erg\,s$^{-1}$ in 0.3-10 keV, respectively). Such a combination of properties suggests the possibility of them being LLIPs \citep{2023MNRAS.523.3192M}.  Follow-up studies should be conducted to confirm this hypothesis.

S-PLUS CV9 and S-PLUS CV10 may be WD photosphere-dominated in the blue.
They may well be WZ Sge-type dwarf novae in quiescence,
whose superoutburst recurrence time is long enough (perhaps decades)
such that they have not been detected in ongoing time-domain surveys. 
Such systems often have the lowest X-ray luminosity of all CVs \citep{2013MNRAS.430.1994R}.  These are therefore arguably the hardest
CVs to discover, except with SED-based surveys such as our work. 

As of now, in addition to validating the proof-of-concept stage of our long-term project to reveal CVs from S-PLUS photometry, we anticipate and encourage further in-depth investigations of the ten CVs reported here. In the subsequent stages of this project, the presented approach will be systematically applied to explore S-PLUS observations to the immediate identification of systems within what we perceive to compose a pure sample. Concurrently, machine learning techniques are being evaluated to enhance the identification of systems toward achieving a more comprehensive sample. 

\subsection{The impact of S-PLUS SEDs for follow-ups of survey discoveries}

In the era of large-scale surveys producing more alerts than the community can
fully digest, astrophysicists of all specialties need a method to quickly select a
subset of interest for follow-up studies.   In this work focusing on CVs, we
have demonstrated that SED-based surveys such as S-PLUS can fill this need
by quickly identifying the nature of newly discovered transients and X-ray sources.
In addition, we have demonstrated that they have the power to discover some interesting objects that have been missed by time-domain and multi-wavelength surveys. 
Therefore, we believe effort such as S-PLUS are
complementary to major projects such as  the LSST \citep{2019ApJ...873..111I} and Zwicky Transient Facility \citep{2019PASP..131a8002B} surveys in the
optical, and e-ROSITA \citep{2021A&A...647A...1P} and Einstein Probe \citep{2015arXiv150607735Y,2025arXiv250107362Y} in X-rays. 
This is because S-PLUS offers an alternative method to standard spectroscopic observations for characterizing still unclassified sources through its photometric SEDs. In fact, the utility of S-PLUS data for stellar characterization has already been demonstrated \citep[e.g.,][]{2021ApJ...912..147W}, and it has also been applied in the extragalactic domain \citep[e.g.,][]{2024MNRAS.531..327N}. 

The coverage of each 1.4 deg x 1.4 deg S-PLUS pointing in twelve bands in 90.8 minutes provides SEDs for over 30-100k targets at once, while reaching depths of $\sim$\,21.5 AB magnitude in the five ugriz broadbands and $\sim$\,20 AB in the seven narrow bands. It brings a real multiplex advantage over traditional spectroscopic follow-ups. 

It is noticeable that a small-aperture telescope, with the right choice of filters, may have such an amazing sinergy with wide-field broad-band surveys in the optical and X-rays as well as pointed observations done with much larger telescopes. 
Similar contributions may come from its twin project, the Javalambre Photometric Local Universe Survey \citep[J-PLUS;][]{2019A&A...622A.176C}, for the Northern Hemisphere, as well as the larger Javalambre-Physics of the Accelerated Universe Astrophysical Survey project \citep[J-PAS;][]{2014arXiv1403.5237B}, also in the Northern Hemisphere, which features 56 photometric bands.

\begin{acknowledgments}
This paper is dedicated to the memory of the good friend and colleague Prof. Dr. João Evangelista Steiner, who devoted his professional life to Astronomy and made significant contributions to the Brazilian astronomical community. 
R.L.O. was partially supported by the Brazilian institutions {\it Conselho Nacional de Desenvolvimento Científico e Tecnológico} (CNPq; PQ-312705/2020-4, PQ-315632/2023-2, and 445047/2024-0) and {\it Fundação de Amparo à Pesquisa do Estado de São
Paulo} (FAPESP; 2020/00457-4). 
A.S.A. was supported by the Brazilian institutions {\it Coordenação de Aperfeiçoamento de Pessoal de Nível Superior} (CAPES; Finance Code 001: 88887.705415/2022-00 and 88887.992353/2024-00), and {\it Fundação Carlos Chagas Filho de Amparo à Pesquisa do Estado do Rio de Janeiro} (FAPERJ; E-26/203.109/2023 - 29438). A.C.K. thanks FAPESP for the support grant 2024/05467-9 and the CNPq. 
R.E. was supported by CAPES (Finance Code 001: 88887.914708/2023-00).
L.A.G-S. acknowledges financial support from {\it Consejo Nacional de Investigaciones Científicas y Técnicas} (CONICET), Agencia I+D+i (PICT 2019–03299), and Universidad Nacional de La Plata (Argentina). M.B.F. acknowledges financial support from CNPq  (307711/2022-6).
We thank Alvaro Alvarez-Candal, Guilherme Limberg, Roberto Saito, Stavros Akras, Swayamtrupta Panda, and Yolanda Jimenez Teja for their suggestions for the manuscript. 
We thank the Swift PI for approving
these observations, and the Swift operations team for implementing them. This research has made use of the SIMBAD database.
\end{acknowledgments}

\vspace{5mm}
\facilities{GAIA, Gemini/GMOS, S-PLUS, Swift(XRT and UVOT)}

\software{\textsc{astropy} \citep{2013A&A...558A..33A,2018AJ....156..123A}, 
\textsc{dragons} \citep{2023RNAAS...7..214L}, 
\textsc{pimms} (https://heasarc.gsfc.nasa.gov/docs/journal/pimms3.html), 
\textsc{xspec} \citep{1996ASPC..101...17A}
}
\bibliography{sample631}{}
\bibliographystyle{aasjournal}

\end{document}